\documentclass[aps,pra,superscriptaddress,twocolumn,nofootinbib]{revtex4-1}
\usepackage{amsmath}
\usepackage{amssymb}
\usepackage{dsfont}
\usepackage{color}
\usepackage[dvipsnames]{xcolor}
\makeatletter
\usepackage{units}
\usepackage{amssymb}
\usepackage{graphicx}
\usepackage{epstopdf}
\usepackage{epsfig}
\usepackage{subfigure}
\usepackage[colorlinks,urlcolor=blue,citecolor=blue]{hyperref}
\makeatother

\usepackage{cancel,ifthen}
\usepackage[normalem]{ulem}

\newcommand{\bref}[3]{{#1}\href{#2}{#3}}

\newcommand{\comment}[2][NoInPuT]{\ifthenelse{\equal{#1}{NoInPuT}}{}{{\color{blue}\sout{#1}}}{\color{red} #2}}
\allowdisplaybreaks
\begin{document}

\title{Dark quantum droplets and solitary waves in beyond-mean-field Bose-Einstein condensate mixtures}

\author{Matthew Edmonds}
\email{m.edmonds@uq.edu.au}
\affiliation{ARC Centre of Excellence in Future Low-Energy Electronics Technologies,
School of Mathematics and Physics, University of Queensland, St Lucia, QLD 4072, Australia}
\affiliation{Department of Physics \& Research and Education Center for Natural Sciences, Keio University, Hiyoshi 4-1-1, Yokohama, Kanagawa 223-8521, Japan}

\begin{abstract}\noindent
Quantum liquid-like states of matter have been realized in an ongoing series of experiments with ultracold Bose gases. Using a combination of analytical and numerical methods we identify the specific criteria for the existence of dark solitons in beyond-mean-field binary condensates, revealing how these excitations exist for both repulsive and attractive interactions, the latter leading to dark quantum droplets with properties intermediate between a dark soliton and a quantum droplet. The phenomenology of the these excitations are explored within the full parameter space of the model, revealing the novel spatial profile of the excitation that differs significantly from the Zakharov-Shabat (ZS) soliton; leading to a negative effective mass that is enhanced in the presence of the quantum fluctuations. Finally the dynamics of pairs of the excitations are explored, showing non-integrable dynamics and dark soliton bound-states in the attractive regime.
%By means of analytical and \me{numerical} methods we identify the specific criteria for the existence of dark solitons in beyond-mean-field condensates, revealing how these excitations exist for both repulsive and attractive interactions, the latter leading to dark quantum droplets with properties intermediate between a dark soliton and a quantum droplet. The dark quantum droplet's physical characteristics are investigated, including calculation of the integrals of motion, revealing their sensitive dependence on physical parameters relevant to the current generation of experiments with quantum gases in the beyond-mean-field limit.
\end{abstract}

\maketitle

\section{Introduction}
Liquid states of matter give rise to a plethora of fluidic phenomena caused by the interaction of atoms with each other, external forces and other matter \cite{cl_book}. For classical fluids, intermolecular potentials give rise to macroscopic consequences such as surface tension and viscosity, as well as transient effects like the Rayleigh-Taylor instability and turbulence, phenomena that can be observed on terrestrial \cite{cengel_book} and astronomical \cite{cc_book} scales. The intrinsic properties of fluids depend critically on their thermodynamic environment, quantum liquids can also share some of the properties of their classical counterparts while also exhibiting unique and unexpected phenomena with no classical analogue \cite{leggett_1999}.

The last few years have seen a series of groundbreaking experiments with degenerate atomic Bose-Einstein condensates which have demonstrated the capacity of these intrinsically weakly correlated systems to manifest liquid-like states of matter in the form of quantum droplets, made from highly magnetic \cite{kadau_2016,barbut_2016,schmitt_2016,chomaz_2016}, mixtures \cite{cabrera_2018,semeghini_2018,cheiney_2018,ferioli_2019,derrico_2019} and also purely Lee-Huang-Yang  \cite{jorgensen_2018,minardi_2019,skov_2021} quantum gases. These surprising discoveries have been attributed to purely quantum mechanical effects in the form of the Lee-Huang-Yang (LHY) correction \cite{lee_1957}, which provides the stabilization required to avoid instability originating from collisional forces.

While there has been intense focus on understanding the ground states of many-body systems, their excitations also play a crucial role in understanding their fundamental behaviour. Recently there has been renewed experimental interest in realizing nonlinear excitations with quantum gases such as dark solitons \cite{aycock_2016,bersano_2018,fritsch_2020,mossman_2022} and domain walls \cite{chai_2021,yao_2022} which provide insight into reduced dimensionality topology in a highly controllable setting. Such states could provide an important resource for future applications in atromtronics \cite{amico_2022} as well as providing fundamental insight into the physics of lower dimensional quantum systems \cite{mistakidis_2022}. 
%\begin{figure}[b]
%\includegraphics[width=1\columnwidth]{figure1.pdf}
%\caption{\label{fig:schem}(color online) Dark quantum droplet schematic. The three-dimensional tubes represent individual condensate mixtures. The grey stripes represent the dark soliton (background) and dark quantum droplet (foreground). The red-white-blue shading indicates the local phase of the excitation.}
%\end{figure}

Quantum gases possessing internal degrees of freedom represent an important testing ground for many body phenomena. These additional degrees of freedom can facilitate unique quantum states that sensitively depend on the nature of the atomic interactions \cite{chin_2010}. The presence of attractive interactions in these systems can ordinarily lead to the collapse of the quantum state; however it was shown theoretically that such a system can in principle be stabilized by beyond-mean-field effects \cite{petrov_2015}. This stimulated an intense interest in the phenomenology of beyond-mean-field physics in these systems -- here fundamental questions such as the role of dimensionality \cite{petrov_2016,zin_2018,llg_2018}, confinement \cite{pathak_2022,debnath_2022}, dynamical \cite{astrakharchik_2018,mithun_2020,saqlain_2022}, collective \cite{tylutki_2020}, coherent \cite{chiquillo_2019} and gauge couplings \cite{tononi_2019}, as well as non-equilibrium \cite{guebli_2021,mithun_2021} effects and phase separation \cite{sturmer_2022} have provided key insight into the unusual liquid-like properties of these ultra-dilute droplets \cite{luo_2021,khan_2022}. Complementary to their existence in degenerate atomic systems, droplet states have also been investigated in other systems such as photonic \cite{wilson_2018}, optomechanical \cite{walker_2022}, as well as in the Helium liquids \cite{barranco_2006}.

While previous works have addressed aspects of the fundamental nature of liquid-like ground states in quantum gases, recent work has focused on investigating the excitations in these systems which possess non-trivial phase windings such as kinks \cite{shukla_2021}, vortices \cite{kartashov_2022}, and dark solitons in dipolar systems \cite{kopycinski_2022}.

The purpose of this work is to study the properties of the excitations in beyond-mean-field Bose-Einstein condensates in one-dimension, including elucidating the fundamental criteria for the existence of dark quantum droplets (DQDs) -- dark soliton-like excitations that exist in the beyond-mean-field model with attractive, rather than repulsive interactions in the cubic-quadratic Schr\"odinger system, as well as characterising their fundamental properties with a complimentary combination of numerical and analytical approaches. %A schematic representation of the dark quantum droplet is presented in Fig.~\ref{fig:schem}. 

The paper is organized beginning with a description of the theoretical model describing the beyond-mean-field Bose-Einstein condensate mixture in Sec.~\ref{sec:bmf}, including the basic solutions and conserved quantities that this model accommodates. The crossover from a dark soliton to the dark quantum droplet is explored in Sec.~\ref{sec:dqd}, including a comparison of the droplets analytical and numerical properties in terms of the droplet's size, integrals of motion and effective mass as well as the dynamics of individual and pairs of dark solitary waves in this system. The paper concludes with a summary and outlook, Sec.~\ref{sec:sum}.    

\section{\label{sec:bmf}Beyond-mean-field model}
The energy of $N=N_{\uparrow}+N_{\downarrow}$ Bose particles with mass $m$ forming a two-component homogeneous atomic Bose-Einstein condensate can be written as
\begin{equation}\label{eqn:bin_en}
E_{\rm 3D}=\int d^3{\bf r}\bigg[\frac{\hbar^2}{2m}\sum_{j}|\nabla\Psi_j({\bf r})|^2+\sum_{j,k}\frac{g_{jk}}{2}n_j({\bf r}) n_k({\bf r})\bigg]
\end{equation}
here $j,k\in\{\uparrow,\downarrow\}$, $g_{jk}=4\pi\hbar^2 a_{jk}/m$ defines the scattering parameter between atoms and $n_{j}({\bf r})\equiv|\Psi_j({\bf r})|^2$ defines the atomic density for component $j$. In order to understand the effect of beyond-mean-field effects, the underlying many body Hamiltonian is diagonalized within the standard Bogoliubov de-Gennes formalism for the weakly interacting limit, after integrating out the transverse spatial degrees of freedom the one-dimensional ground state energy density is obtained as \cite{petrov_2016}
\begin{align}\nonumber
E_{\rm 1D}=&\frac{(\sqrt{g_{\uparrow\uparrow}}n_{\uparrow}{-}\sqrt{g_{\downarrow\downarrow}}n_{\downarrow})^2}{2}{+}g\delta g\frac{(\sqrt{g_{\downarrow\downarrow}}n_{\uparrow}{+}\sqrt{g_{\uparrow\uparrow}}n_{\downarrow})^2}{(g_{\uparrow\uparrow}+g_{\downarrow\downarrow})^2}\\&-\frac{2\sqrt{m}}{3\pi\hbar}(g_{\uparrow\uparrow}n_{\uparrow}+g_{\downarrow\downarrow}n_{\downarrow})^{3/2},\label{eqn:ed1d}
\end{align}
here $g=(g_{\uparrow\uparrow}n_{\uparrow}+g_{\downarrow\downarrow}n_{\uparrow})/n$, $\delta g=g_{\uparrow\downarrow}+\sqrt{g_{\uparrow\uparrow}g_{\downarrow\downarrow}}$ and $n=n_{\uparrow}+n_{\downarrow}$ \cite{zin_2018}. Assuming an equal number of atoms in the spin mixture such that $n_{\uparrow}=n_{\downarrow}\equiv n$ and equal inter-component interaction strengths $g_{\uparrow\uparrow}=g_{\downarrow\downarrow}\equiv g$, Eq.~\eqref{eqn:ed1d} simplifies to $E_{\rm 1D}=\delta g n^2-4\sqrt{2m}(gn)^{3/2}/(3\pi\hbar)$. Then within the local density approximation an effective equation of motion can be derived from the chemical potential $\mu_{\rm QF}[\psi]=\partial E_{\rm 1D}/\partial N$ giving
\begin{equation}\label{eqn:cqschro}
i\hbar\frac{\partial\psi}{\partial t}=\bigg[\frac{\hat{p}_{x}^{2}}{2m}-u\hat{p}_x-\frac{\sqrt{2m}}{\pi\hbar}g^{3/2}|\psi|+\delta g|\psi|^2\bigg]\psi,
\end{equation}
here $u$ defines the excitation's velocity in the moving frame. Equation \eqref{eqn:cqschro} describes the dynamics of the binary system in the equal (miscible) spin limit in the form of a cubic-quadratic nonlinear Schr\"odinger system. Let us consider the fundamental solutions of Eq.~\eqref{eqn:cqschro} in the limits of interest, $g\rightarrow0$ with $\psi(\mu,x\rightarrow\pm\infty)=\pm\sqrt{n_0}$ and for $g\neq0$ with $\psi(\mu,x\rightarrow\pm\infty)=0$. In the first limit the system is integrable with the well known family of {\rm dark soliton} solutions $\psi_{\rm DS}(\mu_{\rm DS},x)=\sqrt{n_0}(\beta\tanh(\beta x/\xi_{\rm DS})+i\sqrt{1-\beta^2})$, where the healing length is $\xi_{\rm DS}=\hbar/\sqrt{mn_0\delta g}$ with velocity $u$, $\beta=\sqrt{1-u^2}$ where $0<u<c$, $c$ is the speed of sound and $n_0=\lim_{x\rightarrow\infty}|\psi(\mu,x)|^2$ defines the constant asymptotic density. Then we consider the second situation where Eq.~\eqref{eqn:cqschro} possesses instead a {\rm quantum droplet} solution \cite{petrov_2016} 
\begin{equation}\label{eqn:qd}
\psi_{\rm QD}(\mu,x)=\frac{\sqrt{n_0}\mu/\mu_{\rm QD}}{1+\sqrt{1-\mu/\mu_{\rm QD}}\cosh(\sqrt{-2\mu m}x/\hbar)}, 
\end{equation}
here the flat-topped droplet state forms as $\mu\rightarrow\mu_{\rm QD}$ where $\mu_{\rm QD}=-4mg^3/9\pi^2\hbar^2\delta g$. We consider the general situation where both $g\neq\delta g\neq0$. As such the model Eq.~\eqref{eqn:cqschro} has a number of important symmetries. From a physical point of view we consider the regularized versions of the atom number, momentum and energy given respectively by
\begin{subequations}\label{eqn:iom}
\begin{align}
&N=\int dx\bigg[n_0 - |\psi(x)|^2\bigg],\label{eqn:an}\\
&P=\frac{i\hbar}{2}\int dx\bigg[\psi\frac{\partial\psi^*}{\partial x}-\psi^*\frac{\partial\psi}{\partial x}\bigg] - \hbar n_0\Delta\phi,\label{eqn:mom}\\ \nonumber
&E=\int dx\bigg[\frac{\hbar^2}{2m}\bigg|\frac{\partial\psi}{\partial x}\bigg|^2+\frac{\delta g}{2}\big(n_0 - |\psi|^2\big)^2 \\&- \frac{2\sqrt{2m}}{3\pi\hbar}g^{3/2}\bigg\{|\psi|^3 - \frac{3}{2}\sqrt{n_0}|\psi|^2 + \frac{1}{2}\sqrt{n_{0}^{3}}\bigg\}\bigg].\label{eqn:edqd}
\end{align}
\end{subequations}
here the phase difference $\Delta\phi=\text{arg}(+\infty)-\text{arg}(-\infty)$. As well as the three integrals of motion Eqs.~\eqref{eqn:iom}, the model Eq.~\eqref{eqn:cqschro} accommodates distinct dilation invariances in the limits $\delta g=0$ and $g=0$. A dilation transformation is a scaling such that $x\rightarrow\sqrt{\alpha}x$ and $t\rightarrow\alpha t$ for $\alpha\in\mathds{R}_{>0}$, and will in general leave a Schr\"odinger system with a single nonlinearity $|\psi|^{n}$ invariant if $\psi(x,t)\rightarrow\alpha^{1/n}\psi(\sqrt{\alpha}x,\alpha t)$. Then, we can see that when $\delta g=0$ the dark soliton solution obeys $\psi_{\rm D}\rightarrow\sqrt{\alpha}\psi_{\rm D}(\sqrt{\alpha}x,\alpha t)$ while for $g=0$ the quantum droplet undergoes the dilation $\psi_{\rm QD}\rightarrow\alpha\psi_{\rm QD}(\sqrt{\alpha}x,\alpha t)$. The competition between the two length scales associated with the interaction parameters $g$ and $\delta g$ facilitates unusual phenomenology in this nonlinear system.
\begin{figure*}[t]
\centering
\includegraphics[width=1.0\textwidth]{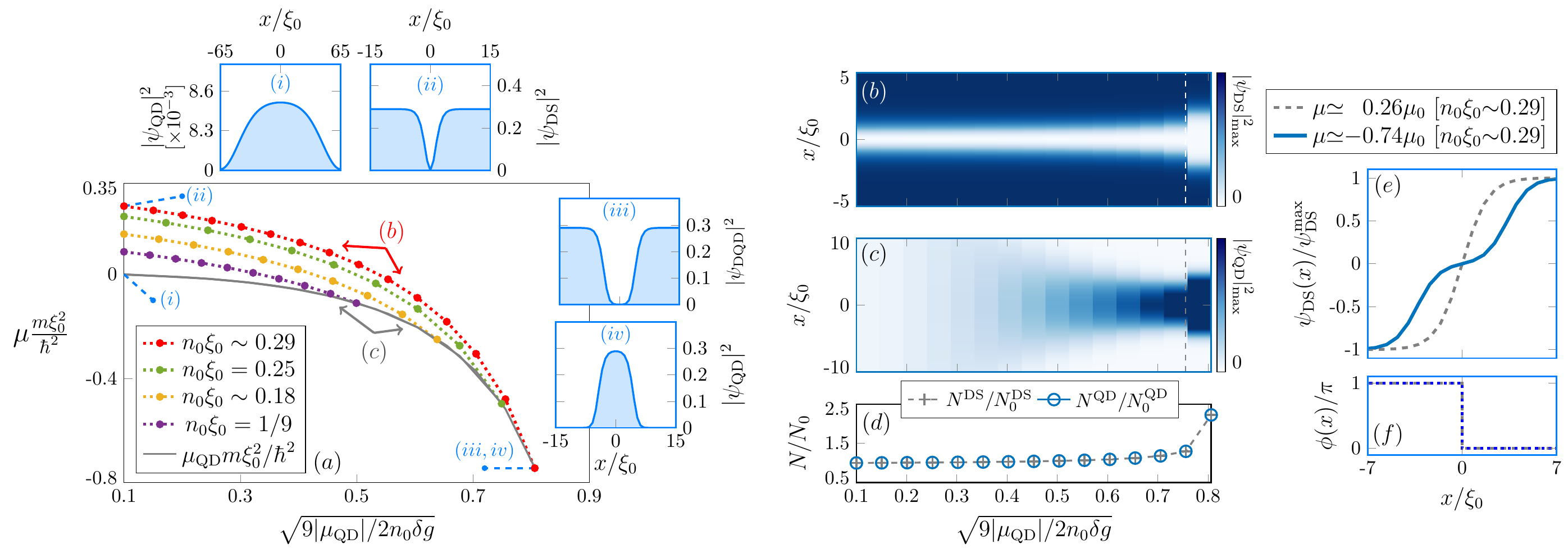}
\caption{\label{fig:trans}(color online) Soliton to dark quantum droplet crossover. The transition from a dark soliton to the dark quantum droplet is shown in panel (a) for the fixed backgrounds $n_{0}\xi_0=0.29,0.25,0.18,1/9$, with the insets (i-iv) showing selected solutions from (a). Heat maps of the excitation and ground state density with $n_{0}\xi_0\sim 0.29$ are shown in (b) and (c) respectively. Panel (d) shows the atom number (Eq.~\eqref{eqn:an}) corresponding to (b) and (c). Comparisons of the excitation's wave function and phase are presented in panels (e) and (f).}
\end{figure*}
From Eq.~\eqref{eqn:cqschro} we can define a set of dimensionless units appropriate for numerical simulations. Since the interactions can be either repulsive ($\mu>0$) or attractive ($\mu<0$) we expect the excitation's size to be of order $\sim\hbar/\sqrt{mn_0\delta g}$ when the mean-field van der Waals term dominates over the LHY term, while in the attractive regime the excitation's size is of order $\sim\hbar/\sqrt{m|\mu_{\rm QD}|}$. Then, the healing length $\xi_0=\hbar/\sqrt{m|\mu_0|}$ with $\mu_0=-\sqrt{2mn_0 g^3}/\pi\hbar+\delta gn_0$ defines the intrinsic length of the system, from this a time scale $\hbar/|\mu_0|$ follows. The resulting dimensionless interaction strength used in the numerical simulations is $\sqrt{9|\mu_{\rm QD}|/2n_0\delta g}$. 

To understand how a particular value of the dimensionless interaction strength changes the sign of the interactions, we consider the homogeneous limit of Eq.~\eqref{eqn:cqschro} where $\psi(x,t)=\sqrt{n_0}\exp(-i\mu_0 t/\hbar)$, then the point at which the interactions change sign is $\mu_0$, which corresponds to $\sqrt{9|\mu_{\rm QD}|/2n_0 \delta g}=\sqrt{\xi_0 n_0}$. Then for $0<\sqrt{9|\mu_{\rm QD}|/2n_0 \delta g}<\sqrt{\xi_0 n_0}$ the interactions are repulsive ($\mu>0$), while for $\sqrt{9|\mu_{\rm QD}|/2n_0\delta g}>\sqrt{\xi_0 n_0}$ we have $\mu<0$, attractive interactions. A comparison of the analytical values of the interaction srength at which $\mu$ changes sign shows close agreement with the numerical data presented in Fig.~\ref{fig:trans}(a).

Using scattering lengths appropriate for $^{\rm 39}$K \cite{derrico_2007} we can compute approximate values for the equilibrium density $n_0$ using \cite{petrov_2016}
\begin{equation}
n_0 = \frac{8}{9\pi^2}\frac{mg^3}{\hbar^2\delta g^2}
\end{equation}
with $m=m_{\rm 39K}=6.5\times 10^{-26}$kg, $a_{\rm\uparrow\uparrow}=a_{\rm\downarrow\downarrow}\simeq 100a_0$, $|a_{\rm\uparrow\downarrow}|\simeq 50a_0$ gives $\delta g\simeq 10^{-37}$Jm and $g=7\times 10^{-38}$Jm. Previous experiments with strongly confined quantum gases have been able to achieve optical confinement with strengths of order $\omega_r\simeq 2\pi\times 10$kHz \cite{haller_2009} which gives a one-dimensional density $n_0\sim 1.6\times 10^4$/m. Admittedly this is quite a small value, but optimistically could be improved in the future with the ever improving toolbox of quantum technologies for cold atom experiments \cite{amico_2022b}. Then, the dimensionless interaction parameter $\sqrt{9|\mu_{\rm QD}|/2n_0\delta g}\sim 1$ similar to the values that will be used in this work.    

\section{\label{sec:dqd}Dark quantum droplets}

\subsection{Dark soliton to dark quantum droplet crossover}
In this section we explore the nature of the solutions to Eq.~\eqref{eqn:cqschro} in the limit $u=0$. Since we are interested in the excited states, we use an iterative (Newton-Raphson) approach to compute these states. An overview of the numerical procedure is given in the Appendix \ref{sec:app}.

We explore the transition from a dark soliton excitation to the dark quantum droplet in Fig.~\ref{fig:trans}. In panel (a) we solve the cubic-quadratic Schr\"odinger equation (Eq.~\eqref{eqn:cqschro}) as a function of the interaction strength for both the excited (dark soliton-like excitation) and quantum droplet ground state. Each dark soliton solution is computed using a Newton-Raphson method with fixed Neumann boundary conditions. From this, the atom number Eq.~\eqref{eqn:an} is calculated. This in turn is used as the input for the ground state quantum droplet's atom number $N_{\rm QD}=\int dx|\psi_{\rm QD}(x)|^2$, so for a fixed value of the interaction strength in Fig.~\ref{fig:trans}(a) $N_{\rm DQD}=N_{\rm QD}$. Each quantum droplet's ground state is computed using an imaginary time Fourier split-step method. The chemical potential is plotted for both situations, and for each fixed boundary condition, it is found that the chemical potential of the dark soliton eventually meets that of the droplet state. We can calculate the critical point at which this occurs by equating the quantum droplet's chemical potential $\mu_{\rm QD}$ with the homogeneous chemical potential $\mu_0$, which leads to the criterion

\begin{equation}\label{eqn:crit}
\sqrt{\frac{9|\mu_{\rm QD}|}{2n_0\delta g}}=(1-10^{-\lambda})\frac{3}{2}\sqrt{n_0\xi_0}
%g^3 = \frac{9\pi^2}{8}\frac{n_0\hbar^2}{m}\delta g^2.
\end{equation}
with corresponding critical chemical potential $\mu_{\rm crit}=-n_0\delta g/2$. Very close to this point, the dark soliton acquires a profile resembling an inverted quantum droplet with a hollow central region, but with an asymmetric wave function. The final simulation point is chosen by including the pre-factor $1-10^{-\lambda}$ in Eq.~\eqref{eqn:crit} with $\lambda=3$ for Fig.~\ref{fig:trans} (for $\lambda\rightarrow\infty$ the size of the droplet diverges). Panels (i-iv) show a number of example density profiles taken from the red-dotted data ($n_{0}\xi_0\sim 0.29$). Far from the transition point at weak attractive (repulsive) values of the chemical potential a broad quantum droplet (narrow dark soliton) is observed (panels (i) and (ii) respectively). Then, very close to the point at which the chemical potentials cross, the dark soliton develops a wide hollow region around its core, while the droplet state at this point becomes narrow and tall (panels (iii) and (iv) respectively). A heat map of the red-dotted transition data from (a) is shown in (b), along with the accompanying quantum droplet ground sate data in panel (c), while the dashed lines in panels (b) and (c) correspond to the solutions (iii) and (iv) discussed above. Following this panels (e) and (f) compare the solutions $\psi_{\rm DS}$, scaled to the asymptotic spatial values and the accompanying phase $\phi(x)=\tan^{-1}({\rm Im}(\psi_{\rm DS})/{\rm Re}(\psi_{\rm DS}))$ respectively for $\mu\simeq 0.32\mu_0$ (dark soliton) and $\mu\simeq -0.74\mu_0$ (dark quantum droplet). Panel (d) presents the atom number $N_{\rm DQD}$ (Eq.~\eqref{eqn:an}) for both situations, showing the gradual increase that occurs as the transition point is approached. Recent work has also studied the existence of dark soliton-like excitations in the binary LHY system (Refs.~\cite{shukla_2021} and \cite{kartashov_2022}). These works interpreted the excitation's unusual shape in terms of a pair of separating kink anti-kink pairs.
\begin{figure}[t]\hspace{-5mm}\centering
\includegraphics[width=0.9\columnwidth]{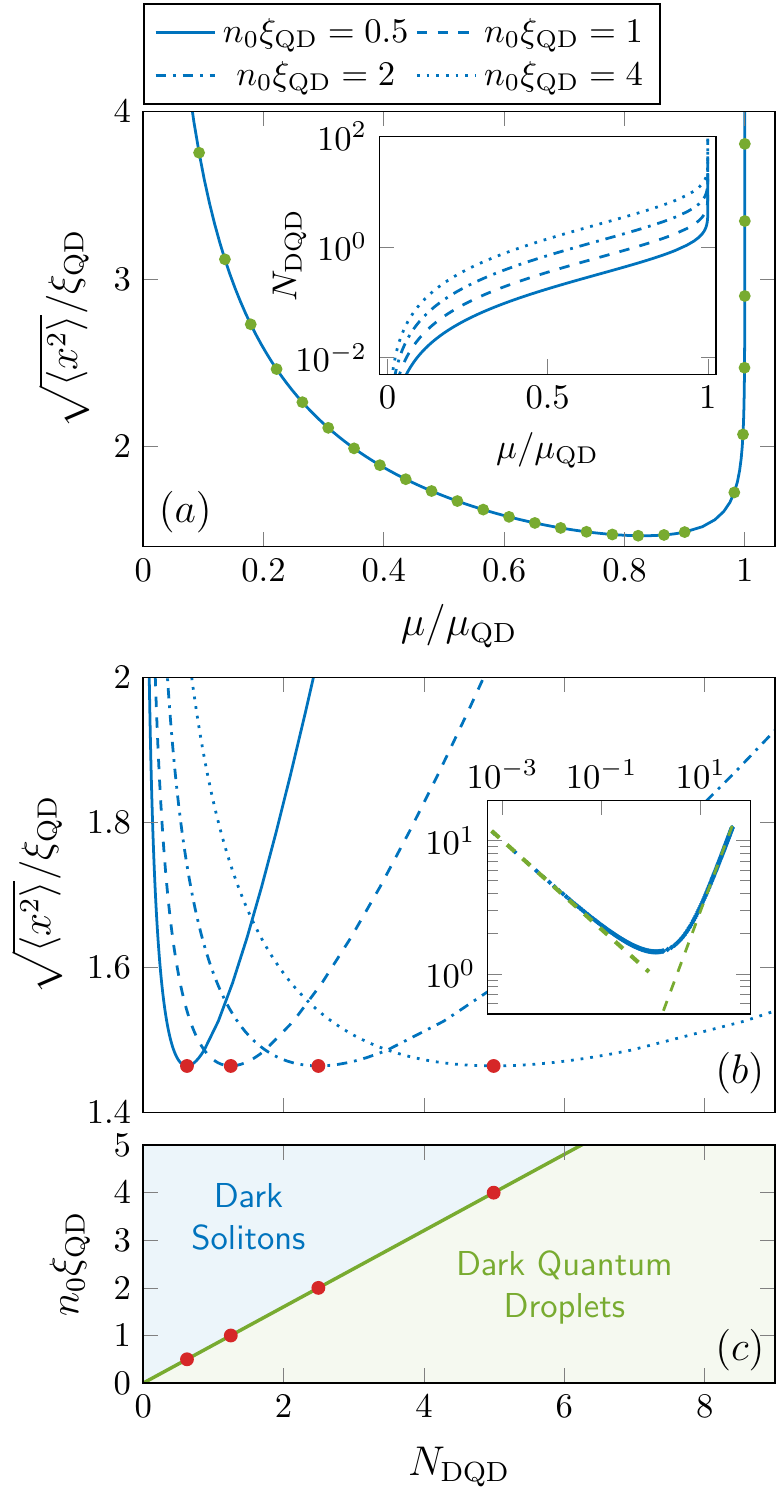}
\caption{\label{fig:xsq}(color online) Dark quantum droplet root-mean-squared width. Panel (a) shows Eq.~\eqref{eqn:xsq}, $\sqrt{\langle x^2\rangle}/\xi_{\rm QD}$ as a function of $\mu/\mu_{\rm QD}$, the minima occurs for $\mu\simeq0.8306\mu_{\rm QD}$, while the inset shows Eq.~\eqref{eqn:ndqd} for several values of $n_0\xi_{\rm QD}$. Panel (b) shows $\sqrt{\langle x^2\rangle}/\xi_{\rm QD}$ instead as a function of $N_{\rm DQD}$, while the inset displays a log-log plot for $n_0\xi_{\rm QD}=\nicefrac{1}{2}$. The minima of $\sqrt{\langle x^2\rangle}/\xi_{\rm QD}$ in (b) are plotted in (c) (green solid) with the red circles correspond to the locations of the four curves individual minima.}
\end{figure}
\subsection{Root-mean-squared width}

The results presented in Fig.~\ref{fig:trans} reveal that as the chemical potential of the dark soliton approaches that of the quantum droplet, the soliton's profile resembled that of an inverted droplet. Previous experimental studies of the soliton to droplet crossover \cite{cheiney_2018,natale_2022} have established that one can define a soliton at relatively small atom numbers, while for large atom numbers a quantum droplet emerges, we can perform an similar distinction here to understand the crossover from a dark soliton to a dark quantum droplet. From Fig.~\ref{fig:trans} (iii-iv) we can infer that

\begin{equation}\label{eqn:lim}
\lim_{\mu\rightarrow\mu_{\rm QD}}\bigg(n_{\rm QD}(\mu,x) + n_{\rm DQD}(\mu,x)\bigg)=n_0.
\end{equation}
Equation \eqref{eqn:lim} will allow us to calculate observables of the dark quantum droplet state. The mean-squared width is an important characteristic which can be used to characterise the behaviour of the dark droplet as the chemical potential approaches that of the quantum droplet's. Similarly to the regularized forms of the atom number, momentum and energy (Eqs.~\eqref{eqn:an}-\eqref{eqn:edqd}) we can also compute the mean-squared width from 

\begin{align}\nonumber
\langle x^2\rangle&=\frac{1}{N_{\rm DQD}(\mu)}\int\limits_{-\infty}^{\infty}dx x^2\bigg[n_0 - \lim_{\mu\rightarrow\mu_{\rm QD}} n_{\rm DQD}(\mu,x)\bigg],\\ \label{eqn:xsq_int}
&{=}\frac{1}{N_{\rm DQD}(\mu)}\int\limits_{-\infty}^{\infty}dx x^2\lim_{\mu\rightarrow\mu_{\rm QD}} n_{\rm QD}(\mu,x)
\end{align}
here the second line, Eq.~\eqref{eqn:xsq_int} has been written using Eq.~\eqref{eqn:lim}. From here the known solution for the droplet $n_{\rm QD}(\mu,x)\equiv|\psi_{\rm QD}(\mu,x)|^2$ (Eq.~\eqref{eqn:qd}) can be used to obtain an expression for both the atom number $N_{\rm DQD}(\mu)$ and the mean-squared width $\langle x^2\rangle$ using the inversion formulae for the polylogarithms for the latter, yielding
\begin{align}\nonumber
&\frac{\langle x^2\rangle}{\xi_{\rm QD}^2}=\frac{N_0}{N_{\rm DQD}(\mu)}\frac{\mu_{\rm QD}}{\mu}\bigg[\frac{1}{3}\bigg(\text{arsech}^{3}\sqrt{1{-}\frac{\mu}{\mu_{\rm QD}}} +\pi^2\times\\&\text{arsech}\sqrt{1{-}\frac{\mu}{\mu_{\rm QD}}}\bigg){-}\sqrt{\frac{\mu}{\mu_{\rm QD}}}\bigg[\text{arsech}^{2}\sqrt{1{-}\frac{\mu}{\mu_{\rm QD}}}{+}\frac{\pi^2}{3}\bigg]\bigg],\label{eqn:xsq}
\end{align}

here $\xi_{\rm QD}=\hbar/\sqrt{m|\mu_{\rm QD}|}$ with the constant $N_0=2n_0\xi_{\rm QD}=\sqrt{2n_{0}^{2}\hbar^2/m|\mu_{\rm QD}}|$. Then the atom number $N_{\rm DQD}(\mu)$ appearing in Eq.~\eqref{eqn:xsq} can be evaluated in a similar manner, giving 
\begin{equation}\label{eqn:ndqd}
\frac{N_{\rm DQD}(\mu)}{N_0}=2\text{artanh}\bigg[\frac{\sqrt{\mu/\mu_{\rm QD}}}{1{+}\sqrt{1{-}\mu/\mu_{\rm QD}}}\bigg]-\sqrt{\frac{\mu}{\mu_{\rm QD}}}.
\end{equation}
Using Eqs.~\eqref{eqn:xsq} and \eqref{eqn:ndqd} we can understand the intrinsic properties of the dark quantum droplet. First, let us derive the asymptotic behaviour of Eqs.~\eqref{eqn:xsq} and \eqref{eqn:ndqd} when $N_{\rm DQD}\gg1$. For the atom number, one finds the relationship between the chemical potential and $N_{\rm DQD}$ is $N_{\rm DQD}(\mu)/N_0=\ln(2/\sqrt{1-\mu/\mu_{\rm QD}})$. Hence the atom number $N_{\rm DQD}(\mu)$ diverges logarithmically as $\mu\rightarrow\mu_{\rm QD}$. Expanding Eq.~\eqref{eqn:xsq} for $\mu\rightarrow\mu_{\rm QD}$ and using the asymptotic form of Eq.~\eqref{eqn:ndqd}, the root-mean square width in the limit $N_{\rm DQD}\gg1$ is
\begin{equation}\label{eqn:asdqd}
\frac{\sqrt{\langle x^2\rangle}}{\xi_{\rm QD}}\underset{\mu\rightarrow\mu_{\rm QD}}{=}\frac{N_{\rm DQD}}{\sqrt{3}N_0},
\end{equation} 
showing that the effective width of the dark quantum droplet diverges linearly in a fashion qualitatively similar to the quantum droplet \cite{astrakharchik_2018}. 
\begin{figure}[b]
\includegraphics[width=0.95\columnwidth]{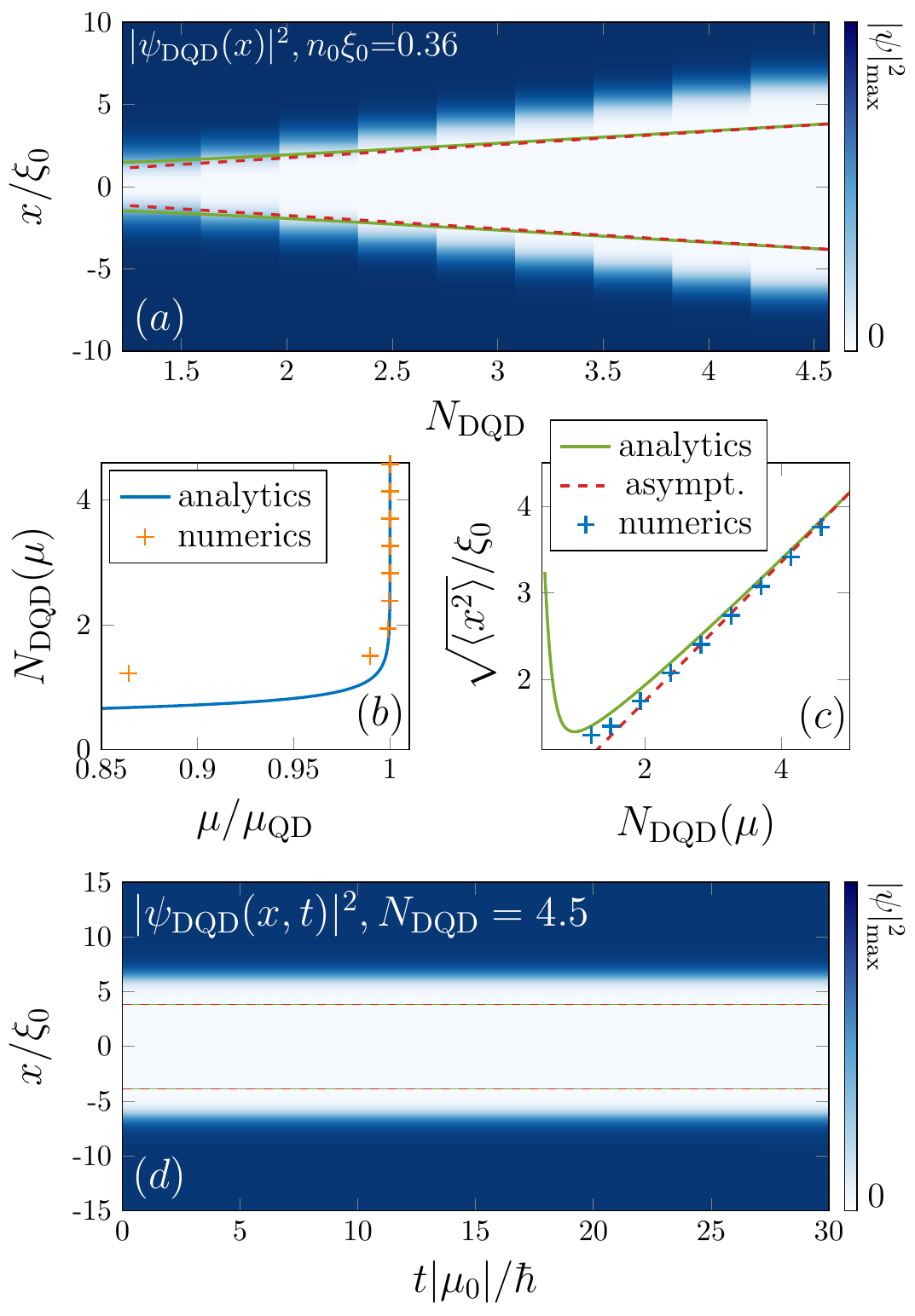}
\caption{\label{fig:wdiv}(color online) Dark quantum droplet width comparison. Numerical solutions for the density $|\psi(x)|^2$ to Eq.~\eqref{eqn:cqschro} are shown in (a) for $n_0\xi_0=0.36$, with the green solid and red dashed lines computed from Eqs.~\eqref{eqn:xsq} and \eqref{eqn:asdqd}. The atom number $N_{\rm DQD}(\mu)$ as a function of the chemical potential $\mu/\mu_{\rm QD}$ is shown in (b) for the analytical result Eq.~\eqref{eqn:ndqd} (solid blue) and numerical values (orange pluses), while (c) compares the root-mean-squared width $\sqrt{\langle x^2\rangle}/\xi_0$ computed using Eq.~\eqref{eqn:xsq} (solid green), Eq.~\eqref{eqn:asdqd} (dashed red) and from the numerical values (blue pluses). Panel (d) shows the real time propagation for $t_{\rm tot}=30\hbar/|\mu_{0}|$ of the final dataset from (a), here $N_{\rm DQD}\sim 4.5$ ($\lambda=9$).}
\end{figure}
Figure \ref{fig:xsq} shows the root-mean-squared width of the dark quantum droplet, Eq.~\eqref{eqn:xsq}. Panel (a) shows the behaviour of the width $\sqrt{\langle x^2\rangle}/\xi_{\rm QD}$ as a function of the chemical potential $\mu/\mu_{\rm QD}$, while the inset shows the atom number Eq.~\eqref{eqn:ndqd} for several values of the background density $n_0\xi_{\rm QD}=\nicefrac{1}{2},1,2,4$; increasing $n_0\xi_{\rm QD}$ has the effect of giving an overall increase to $N_{\rm DQD}$. Note that there is no dependency of $\langle x^2\rangle/\xi_{\rm QD}^2$ on the background density $n_0\xi_{\rm QD}$ when plotted as a function of the dimensionless chemical potential $\mu/\mu_{\rm QD}$. The second panel (b) shows the root-mean-squared width $\sqrt{\langle x^2\rangle}/\xi_{\rm QD}$ as a function of the atom number, for the same values of background density shown in the inset of (a). Increasing $n_0\xi_{\rm QD}$ has the effect of stretching $\sqrt{\langle x^2\rangle}/\xi_{\rm QD}$ such that the linear part ($N_{\rm DQD}(\mu)\gg1$) associated with the dark quantum droplet occurs at larger values of $N_{\rm DQD}(\mu)$. The minima of $\sqrt{\langle x^2\rangle}/\xi_{\rm QD}$ also shift to larger values of $N_{\rm DQD}(\mu)$ as $n_0\xi_{\rm QD}$ is increased. The inset of Fig.~\ref{fig:xsq}(b) shows the dataset for $n_0\xi_{\rm QD}=\nicefrac{1}{2}$ in (b) in a log-log plot. The dashed lines show the asymptotic forms of Eq.~\eqref{eqn:xsq} for $\mu/\mu_{\rm QD}\rightarrow 0$, $\sqrt{\langle x^2\rangle}\sim N_{\rm DQD}^{-1/3}$ and $\mu/\mu_{\rm QD}\rightarrow 1$, $\sqrt{\langle x^2\rangle}/\xi_{\rm QD}=N_{\rm DQD}/(\sqrt{3}N_0)$ \cite{qd_comment}, the second of these limits being appropriate to the dark quantum droplet. The minima of $\sqrt{\langle x^2\rangle}/\xi_{\rm QD}$ are plotted in (c) as a function of $N_{\rm DQD}$ (green solid) with the red circles corresponding to the locations of the four curves minima in (b). The shaded blue and green regions indicate the parameter regimes where we expect dark solitons and dark quantum droplets respectively. 

A comparison of the dark quantum droplet's analytical atom number and root-mean-squared width with the numerically obtained values is explored next in Fig.~\ref{fig:wdiv}. Stationary state solutions to Eq.~\eqref{eqn:cqschro} are shown in (a) for $n_0\xi_0=0.36$. The interaction strength is chosen using Eq.~\eqref{eqn:crit} again using the pre-factor $1-10^{-\lambda}$ with $\lambda=1,2,\dots,9$. The analytic atom number of Eq.~\eqref{eqn:ndqd} (solid blue) is plotted along with the equivalent values computed from the numerical (orange pluses) data in panel (b), here good agreement is found as $\mu\rightarrow\mu_{\rm QD}$. The root-mean-squared width $\sqrt{\langle x^2\rangle}/\xi_0$ is compared from Eq.~\eqref{eqn:xsq} (solid green) and the numerical data (blue pluses). The agreement is found to improve as $\lambda$ increases, and it was found that due to the underlying logarithmic divergence of $N_{\rm DQD}(\mu)$ as $\mu\rightarrow\mu_{\rm QD}$ obtaining a convergence between the analytical and numerical results in general requires very large $\lambda$, which becomes impractical for numerical simulations, but could be an interesting question to explore in a future experiment. The green solid and red dashed lines in (a) are computed from Eqs.~\eqref{eqn:xsq} and \eqref{eqn:asdqd} respectively. Panel (d) shows the dynamics of the $\lambda=9$ solution, showing the stationary profile of the excitation. The green and red-dashed lines are plots of Eqs.~\eqref{eqn:xsq} and \eqref{eqn:asdqd} respectively.  
\begin{figure}[t]
\includegraphics[width=0.945\columnwidth]{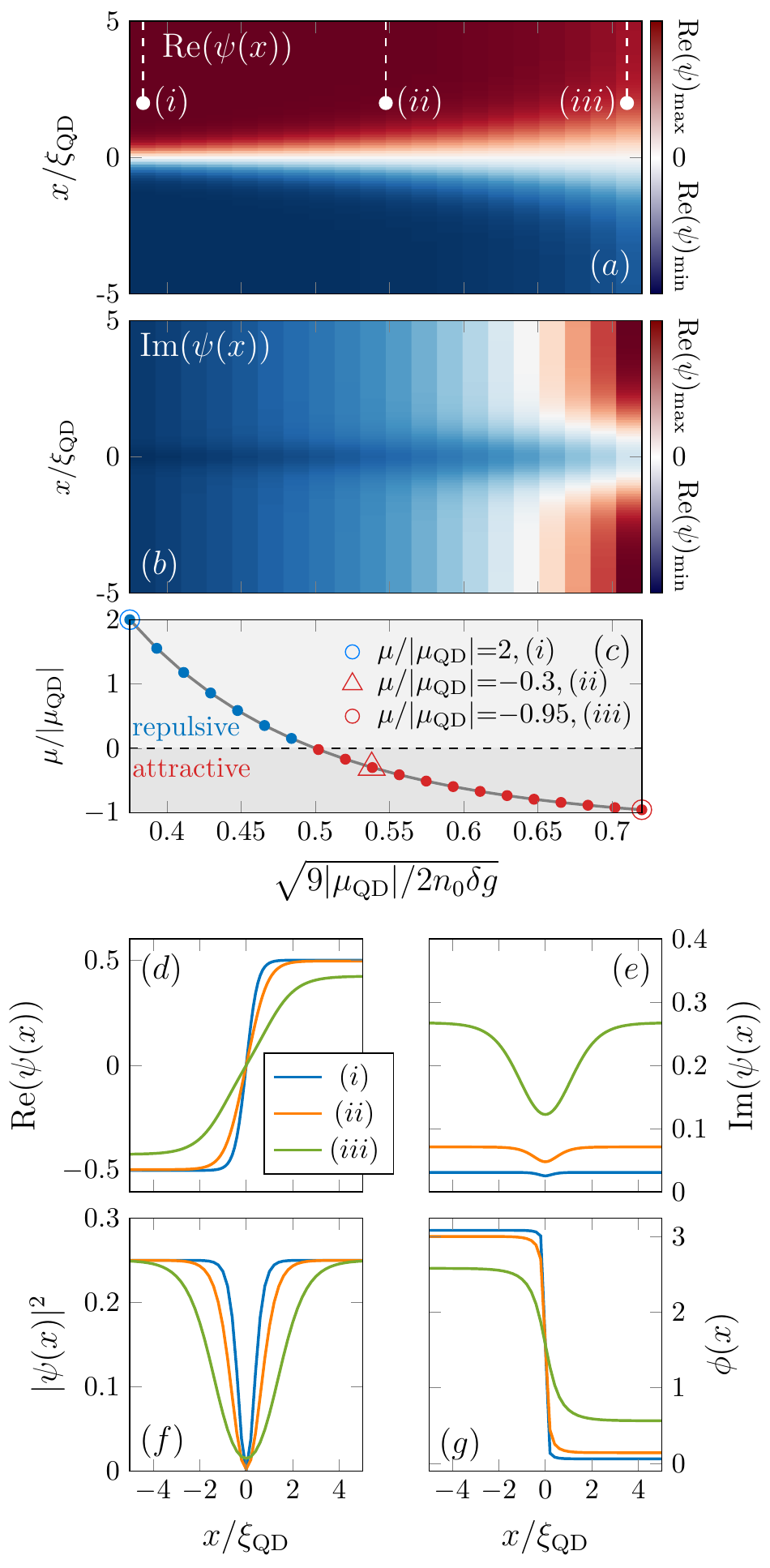}
\caption{\label{fig:mframe} (color online) Moving frame solutions. The real and imaginary parts of the solutions to Eq.~\eqref{eqn:cqschro} are presented as a function of $\sqrt{9|\mu_{\rm QD}|/2n_0\delta g}$ respectively in panels (a) and (b). The chemical potential is shown as a function of the interaction strength in (c), with the individual highlighted points (i)-(iii). Individual solutions are shown in (d) ($\text{Re}(\psi)$) and (e) ($\text{Im}(\psi)$) for $\sqrt{9|\mu_{\rm QD}|/2n_0\delta g}=\{0.375,0.538,0.72\}$, while (f) and (g) show the same data in terms of density $|\psi(x)|^2$ and phase $\phi(x)$ corresponding to (i), (ii) and (iii) respectively.}
\end{figure}
\subsection{Moving frame solutions and integrals of motion}
\begin{figure}[t]
\includegraphics[width=\columnwidth]{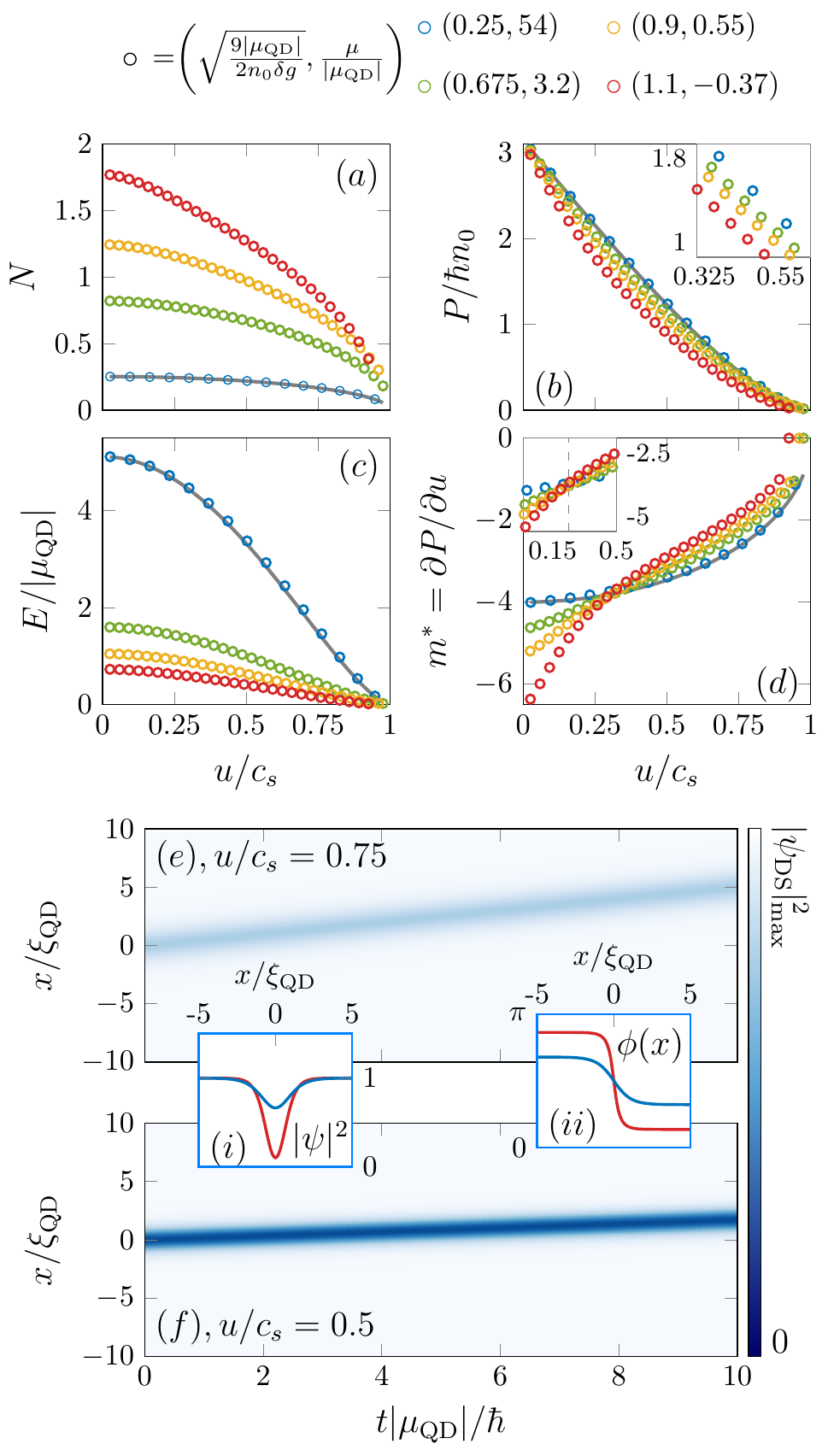}
\caption{\label{fig:iom} (color online) Integrals of motion and dynamics. The four panels (a)-(d) show the calculated values of the quantities $N,P,E$ and $m^*$ given by Eqs.~\eqref{eqn:an}-\eqref{eqn:edqd}, and \eqref{eqn:em}, the key (top) indicates the values of the interaction strength $\sqrt{9|\mu_{\rm QD}|/2n_0\delta g}$ and chemical potential $\mu/|\mu_{\rm QD}|$ that each dataset corresponds to. Panels (e) and (f) show example dynamics taken from the red data for $u/c_s=0.75,0.25$ respectively. Inset (i) and (ii) display the initial density $|\psi(x,0)|^2$ and phase $\phi(x)$ from (e) and (f).}
\end{figure}

In this subsection we investigate the solutions to the beyond mean-field model in the moving frame such that $u\neq 0$ where $u$ is the velocity in the moving frame and $\hat{H}_{\rm bMF}$ denotes the beyond mean-field Hamiltonian appearing in Eq.~\eqref{eqn:cqschro}. In the limit that $g=0$ we recover the well-known Zakharov-Shabat (ZS) solution \cite{zakharov_1973}, here the allowed solutions exist of the interval $0\leq u <c_s$ where $c_s$ is the speed of sound. The depth of the excitation is directly related to it's velocity through $\sqrt{n_{\rm min}/n_0}=u/c_s$ where $n_{\rm min}$ is the density at the centre of the phase twist and $n_0$ is the background, hence the faster the excitation moves the smaller it's depth.
\begin{figure*}[t]
\includegraphics[scale=0.825]{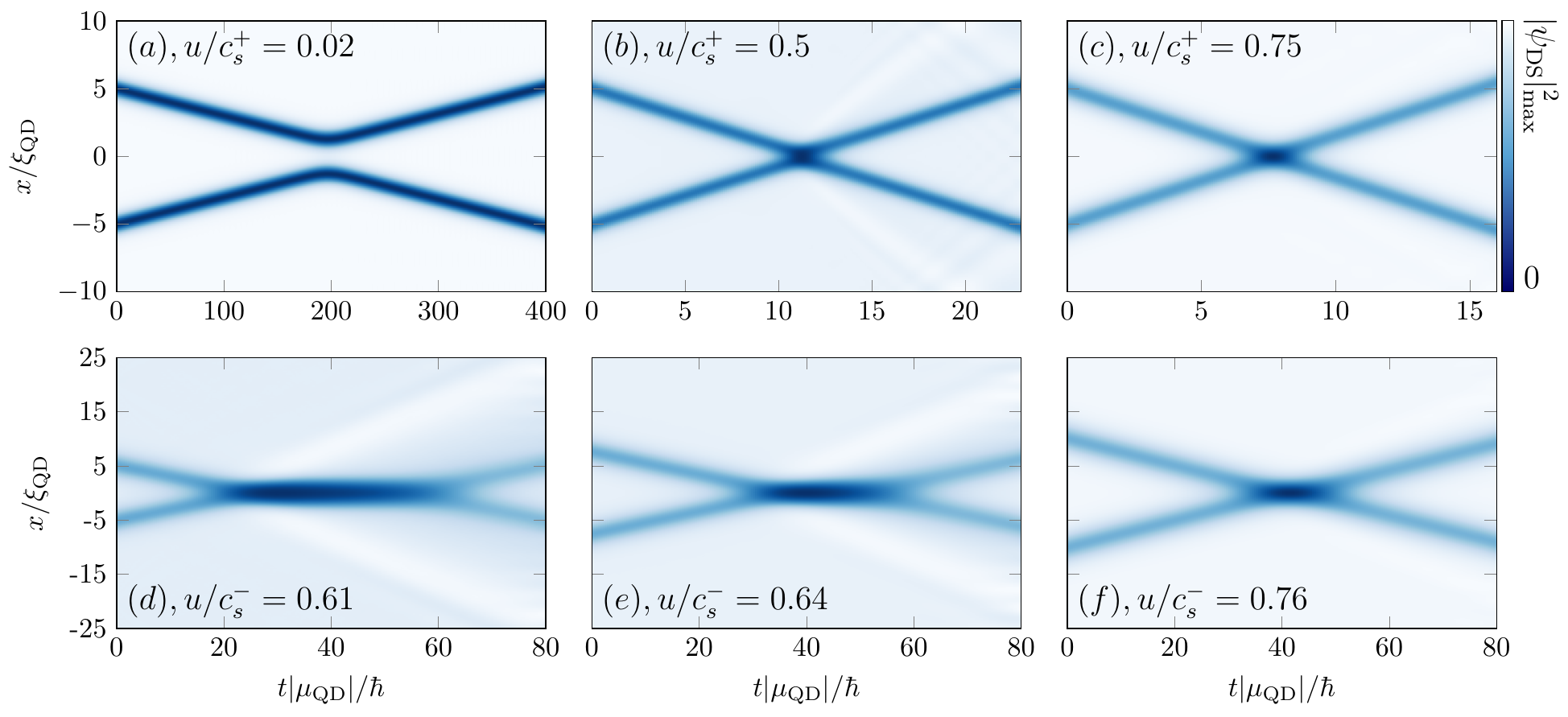}
\caption{\label{fig:dyn} (color online) Excitation collision dynamics. Initial Collisional states are prepared according to Eq.~\eqref{eqn:coll} for fixed background $n_0\xi_{\rm QD}=0.5$. Panels (a)-(c) show collisions for repulsive interactions with $(\sqrt{9|\mu_{\rm QD}|/2n_0\delta g},\mu/\mu_{\rm QD})=(0.63,0.55)$, and speed of sound $c_{s}^{+}\simeq 1.74\sqrt{|\mu_{\rm QD}|/m}$. Panels (d)-(f) show collisions for attractive interactions with $(\sqrt{9|\mu_{\rm QD}|/2n_0\delta g},\mu/\mu_{\rm QD})=(1.1,-1.03)$, and speed of sound $c_{s}^{+}\simeq 0.66\sqrt{|\mu_{\rm QD}|/m}$.}
\end{figure*} 
Figure \ref{fig:mframe} presents the solutions to Eq.~\eqref{eqn:cqschro} as the dimensionless interaction strength is varied for a fixed background density $n_0\xi_{\rm QD}=0.25$ and velocity $u=0.1\sqrt{|\mu_{\rm QD}|/m}$. Panels (a) and (b) show the real $\text{Re}(\psi)$ and imaginary $\text{Im}(\psi)$ parts of the wave function. The top panel shows that the width of the real part increases with interaction strength, the imaginary part shows a marked departure from the ZS solution, whose imaginary part, $\text{Im}(\psi)=\sqrt{n_0}u$ is a constant. We observe that the spatial structure of $\text{Im}(\psi)$ develops a minima localized at the centre of the phae twist as the interaction strength increases. Panel (c) shows the chemical potential as a function of the interaction strength, with the three solutions (i-iii) highlighted. Then, the spatial structure of Im($\psi$) can be seen clearly in panel (e)  where $\text{Im}(\psi)$ evolves from an almost constant solution (blue data (i), $\sqrt{9|\mu_{\rm QD}|/2n_0\delta g}=0.375$) to one with a pronounced minimum (green data (iii), $\sqrt{9|\mu_{\rm QD}|/2n_0\delta g}=0.72$). This behaviour is attributed to the presence of the quantum fluctuations. The density $|\psi|^2$ and phase data $\phi(x)$ corresponding to (d) and (e) are shown in (f) and (g). 

Next we consider the effect varying the excitation's velocity $u$. To understand the role that quantum fluctuations play we can compute the three integrals of motion given by Eqs.~\eqref{eqn:an}-\eqref{eqn:edqd}. In the limit that $g\rightarrow 0$ these quantities can be computed exactly in analytical form, and are given by
\begin{subequations}
\begin{align}
N_{\delta g} &= 2\xi_{\delta g}n_0\beta,\\
P_{\delta g} &= -\frac{2\hbar n_0u\beta}{c_{\delta g}} + 2\hbar n_0\text{arctan}\bigg(\frac{c_{\delta g}\beta}{u}\bigg),\\
E_{\delta g} &= \frac{4}{3}n_0\hbar c_{\delta g}\beta^3.
\end{align}
\end{subequations}
Here one has $\xi_{\delta g}=\hbar/\sqrt{mn_0\delta g}$ and $c_{\delta g}=\sqrt{n_0\delta g/m}$. Then one additional quantity can be obtained using the momentum $P_{\delta g}$, the excitations effective mass 
\begin{equation}\label{eqn:em}
m^{*}=\frac{\partial P}{\partial u}
\end{equation}
which is given by $m^{*}=-4\hbar n_0\beta/c_{\delta g}$. 

These four quantities are computed from Eqs.~\eqref{eqn:an}-\eqref{eqn:edqd}, and presented in panels (a)-(d) of Fig.~\ref{fig:iom} for several fixed interaction strengths ranging from repulsive $(\sqrt{9|\mu_{\rm QD}|/2n_0\delta g},\mu/|\mu_{\rm QD}|)=(0.25,54)$ to attractive $(\sqrt{9|\mu_{\rm QD}|/2n_0\delta g},\mu/|\mu_{\rm QD}|)=(1.1,-0.37)$. The atom number $N$ presented in (a) shows how the size of the excitation's core increases as the interaction strength is increased for a given velocity, eventually approaching zero as the speed of sound is reached. The momentum $P$ is presented in (b). This quantity has a maximum value for $u=0$ of $P/\hbar n_0=\pi$, decreasing to zero as the speed of sound is reached. Increasing the interaction strength has the effect of 'bending' this quantity downwards (see inset). Next, the regularized energy $E$ is computed in (c). For a given velocity $u$, the interaction strength determines the curve with the largest energy, here the gap between the blue ($\mu/|\mu_{\rm QD}|=54$) and green ($\mu/|\mu_{\rm QD}|=3.2$) is caused by the large reduction in repulsive energy. The effective mass $m^{*}$, Eq.~\eqref{eqn:em} is presented in panel (d), here the four datasets cross for $u/c_s\sim 0.32$. Then at velocities approaching $u=0$ the effect of the quantum fluctuations increasingly cause this quantity to have a larger negative value (red data), which could be probed in a future experiment by measuring the oscillation frequency of a beyond-mean-field dark soliton in a harmonic trap \cite{becker_2008}. For $0.32\lesssim u/c_s<1$ the quantum fluctuations instead enhance $m^{*}$ for increasing interaction strength.

The final two panels of Fig.~\ref{fig:iom} show example space-time dynamics for data taken from the red points in panels (a)-(d), here $u/c_s=0.75$ and $0.75$ for (e) and (f) respectively. The scale of both heat maps are the same to highlight the different depths of the excitations. The insets show the initial density $|\psi(x,0)|^2$ and phase $\phi(x)$ for (e) (blue data) and (f) (red data).

\subsection{Excitation collision dynamics}      
The dark solitary wave-like excitation's static properties as presented in the previous sections are useful as a measure of their fundamental properties, however a more comprehensive understanding of their behaviour naturally incorporates their collision dynamics. We simulate the dynamics of pairs of moving frame dark solitary wave-like solutions to the beyond mean-field model of Eq.~\eqref{eqn:cqschro}, in particular we consider a symmetric initial state of the form
\begin{equation}\label{eqn:coll}
\psi(x,t_0) = \psi_{+}(x-x_0,-v,\mu) + \psi_{-}(x+x_0,+v,\mu)
\end{equation}
where $x_0$ defines the initial centre of mass of the solution with velocity $v$ and chemical potential $\mu$. Figure \ref{fig:dyn} presents simulations of the collisions with a fixed background density of $n_0\xi_{\rm QD}=0.5$ for repulsive and attractive interaction strengths. Panels (a)-(c) show collisions with weak repulsive interactions $\sqrt{9|\mu_{\rm QD}|/2n_0\delta g}=0.63$ with $\mu/\mu_{\rm QD}=0.55$ and $x_0/\xi_{\rm QD}=5$. Panel (a) shows a slow collision between two excitations, here $u/c_{s}^{+}=0.02$ ($c_{s}^{\pm}$ represents the speed of sound for repulsive $c_{s}^{+}\simeq 1.74\sqrt{|\mu_{\rm QD}|/m}$ and attractive $c_{s}^{-}\simeq 0.66\sqrt{|\mu_{\rm QD}|/m}$ interactions) and there is an effective repulsion at at the collision point \cite{weller_2008,theocharis_2010}. For $u/c_{s}^{+}=0.5$ the excitations instead pass through each other, with a small amount of sound emission attributed to the proximity of these parameters to the attractive region of the parameter space, i.e. the effect of quantum fluctuations. Then panel (c) shows a faster collision with $u/c_{s}^{+}=0.75$.

Next we simulate collisions with attractive interactions, panels (d)-(f). Here $\sqrt{9|\mu_{\rm QD}|/2n_0\delta g}=1.1$ and $\mu/\mu_{\rm QD}=-1.03$ and $x_0/\xi_{\rm QD}=10,7.5,5$ for (d)-(f) respectively. The excitation's dynamics are found to be qualitatively different to the case of repulsive interactions. In (d) and (e) we observe the formation of short-lived bound states of pairs of excitations. The existence of bound pairs of excitations is attributed to the balance of attractive and repulsive forces in Eq.~\eqref{eqn:cqschro}. Below a critical velocity, the net attractive nonlinear interactions can accommodate a molecule-like state \cite{campbell_1986}. The length of the bound state $t_{\rm bs}$ depends on the excitation's initial velocity - for $u/c_{s}^{-}=0.61$ we find $t_{\rm tb}\simeq 50\hbar/|\mu_{\rm QD}|$, and for $u/c_{s}^{-}=0.64$ we find $t_{\rm tb}\simeq 30\hbar/|\mu_{\rm QD}|$ instead. The emission of radiation in the form of sound is observed in both cases, contributing to the eventual breaking of the bound states. Finally panel (f) shows a faster collision with $u/c_{s}^{-}=0.76$ showing a quasi-elastic collision, with a reduced amount of sound emission. Animations of the solitary waves dynamics corresponding to the data in panels Fig.~\ref{fig:dyn}(c) and (e) are included as supplementary material \cite{supp}.

\section{\label{sec:sum}Summary}
In this work we have explored the phenomenology of dark quantum droplets and solitary waves, revealing the criteria for the existence of dark quantum droplets in beyond-mean-filed Bose-Einstein condensate mixtures. The crossover from the dark soliton at weak repulsive to dark quantum droplets at attractive interaction strengths was found to depend sensitively on the interaction strength, a situation that was explored by comparing the analytical and numerical values of the excitation's root-mean-squared width, finding improving agreement as the transition point is approached. We then explored the beyond-mean-field solutions at finite velocity, revealing the departure of the excitation's shape from the Zakharov-Shabat solution. The integrals of motion of the exciation were computed, allowing the calculation of the excitation's effective mass, which was found to be strongly affected by the quantum fluctuations. Finally, the dynamics of pairs of the dark solitary waves were explored, revealing the existence of bound states in the attractive regime.

Due to the unusual profile of the dark quantum droplet, they could find useful application for example hosting qubits similar to proposals for dark solitons \cite{shaukat_2017}, as well as for matter-wave box traps \cite{navon_2021}, providing an alternate route to realising matter-wave traps in a controllable environment.

For future studies and given the results presented in this work concerning the excitation's effective mass, it would be intriguing to understand the behaviour of the dark quantum droplet in a harmonic trap, and how their oscillation frequency depends on the properties of the excitation \cite{busch_2000}. The dynamical behaviour, such as constructing Toda-like lattices provides another future direction \cite{ma_2016}.     

\section{Acknowledgements}
I thank Thomas Bland for helpful comments on the manuscript. This research was supported by the Australian Research Council Centre of Excellence in Future Low-Energy Electronics Technologies (Project No. CE170100039) and funded by the Australian government, and by the Japan Society of Promotion of Science Grant-in-Aid for Scientific Research (KAKENHI Grant No. JP20K14376).

\appendix*
\section{\label{sec:app}Newton-Raphson method}
Here we give an overview of the numerical method used to procure the dark quantum droplet solutions to the cubic-quadratic Schr\"odinger equation (Eq.~\eqref{eqn:cqschro} of the text). This type of approach has been used previously to study excitations in superfluid systems such as vortices \cite{winiecki_1999}, solitons in dipolar \cite{bland_2015,edmonds_2016} and magnetic systems \cite{chai_2022}. Our system differs from previous studies due to the presence of mixed nonlinearities. We consider a general scheme at finite velocity. First we write a function whose solutions we seek in the the Galilean-boosted frame as
\begin{equation}\label{eqn:fz}
F[\psi] = \bigg(\hat{H}_{\rm cqGPE} - u\hat{p}_x -\mu\bigg)\psi,
\end{equation}
where $u$ is the excitations velocity. Then Eq.~\eqref{eqn:fz} can be translated into the iterative scheme
\begin{equation}\label{eqn:fza}
F_{u}(\psi^{p+1})\approx F(\psi^{p}) + \sum_{v=1}^{N}\mathcal{J}_{u,v}\delta\psi_{v}\approx0,
\end{equation}
here $\delta\psi=\psi^{p+1}-\psi^{p}$ and $\mathcal{J}_{u,v}$ defines the matrix elements of the Jacobian. The solutions to Eq.~\eqref{eqn:fz} are in general complex valued, and since Newton-Raphson methods only work with real data we write the discrete $\psi(x)$ comprising $N$ complex numbers as $2N$ real numbers such that $\text{Re}(\psi(x))\equiv \psi_{j,0}$ and $\text{Im}(\psi(x))\equiv \psi_{j,1}$, the second subscript referring to the real and imaginary components. Then one can write the discrete form of Eq.~\eqref{eqn:fz} as
\begin{widetext}
\begin{align}\nonumber
f_{j,s}{=}&{-}\frac{\hbar^2}{2m}\bigg[\frac{\psi_{j-1,s}-2\psi_{j,s}+\psi_{j+1,s}}{\Delta x^2}\bigg]+(2s-1)\hbar u\bigg[\frac{\psi_{j+1,1-s}-\psi_{j-1,1-s}}{2\Delta x}\bigg]+\bigg\{{-}\frac{\sqrt{2m}}{\pi\hbar}g^{3/2}\sqrt{\psi_{j,0}^{2}{+}\psi_{j,1}^{2}}\\&{+}\delta g(\psi_{j,0}^{2}{+}\psi_{j,1}^{2})\bigg\}\psi_{j,s}-\mu\psi_{j,s}.
\end{align}
\end{widetext}
The boundary conditions for the problem are treated as the Neumann type, such that
\begin{equation}
\frac{d\psi}{dx}\bigg|_{x=\pm L}=0,
\end{equation}
which translates into taking $\psi_{1,s}-\psi_{0,s}=0$ and $\psi_{N+1,s}-\psi_{N,s}=0$ for the kinetic term and $\psi_{2,s}-\psi_{0,s}=0$ and $\psi_{N+1}-\psi_{N-1,s}=0$ for the momentum operator. The matrix elements of the Jacobian appearing in Eq.~\eqref{eqn:fza} are found from $\mathcal{J}_{k,r}^{j,s}=\partial f_{j,s}/\partial\psi_{k,r}$. Using $\partial\psi_{j,s}/\partial\psi_{k,r}=\delta_{j,k}\delta_{s,r}$ we obtain
\begin{widetext}
\begin{align}\nonumber
\mathcal{J}_{k,r}^{j,s}=-&\frac{\hbar^2}{2m}\delta_{s,r}\bigg[\frac{\delta_{k,j-1}-2\delta_{k,j}+\delta_{k,j+1}}{\Delta x^2}\bigg]+(2s-1)\delta_{1-s,r}\hbar u\bigg[\frac{\delta_{k,j+1}-\delta_{k,j-1}}{2\Delta x}\bigg]{-}\frac{\sqrt{2m}}{\pi\hbar}g^{3/2}\delta_{k,j}\bigg[\psi_{j,s}\frac{\delta_{0,r}\psi_{j,0}{+}\delta_{1,r}\psi_{j,1}}{\sqrt{\psi_{j,0}^{2}{+}\psi_{j,1}^{2}}}\\{+}&\delta_{r,s}\sqrt{\psi_{j,0}^{2}{+}\psi_{j,1}^{2}}\bigg]+\delta g\delta_{k,j}\bigg[2\psi_{j,s}(\delta_{0,r}\psi_{j,0}{+}\delta_{1,r}\psi_{j,1}){+}\delta_{r,s}(\psi_{j,0}^{2}{+}\psi_{j,1}^{2})\bigg]-\mu\delta_{k,j}\delta_{s,r},\label{eqn:jcbn}
\end{align}
\end{widetext}
which defines a $2N\times2N$ matrix. Then stationary solutions can be obtained to Eq.~\eqref{eqn:cqschro} using Eqs.~\eqref{eqn:fz}-\eqref{eqn:jcbn} using a tolerance based approach for $\delta\psi$. As such we employ the Frobenius norm $||\delta\psi||=\big(\sum_{j=1}^{2N}\delta|\psi_{j}|^{2}\big)^{1/2}$ as a measure which is deemed convergence after falling below a predefined value, typically $||\delta\psi||<10^{-10}$. Computation of $\delta\psi$ at each step is accomplished by solving the linear system $\boldsymbol{\mathcal{J}}\delta\psi=-\boldsymbol{F}$ using a stabilized biconjugate gradient method which exploits the symmetry of the Jacobian to expedite the solution of the linear system by avoiding matrix inversion. The Newton-Raphson method requires an initial guess for $\psi(x)$, which we take as the dark soliton solution to the cubic Schr\"odinger equation. An example Python script for generating a dark quantum droplet can be found here \cite{python_code}.

\end{document}